\documentclass[sigconf]{acmart}

\AtBeginDocument{%
  }

\copyrightyear{2026}
\acmYear{2026}
\setcopyright{cc}
\setcctype{by}
\acmConference[ICSE '26]{2026 IEEE/ACM 48th International Conference on Software Engineering}{April 12--18, 2026}{Rio de Janeiro, Brazil}
\acmBooktitle{2026 IEEE/ACM 48th International Conference on Software Engineering (ICSE '26), April 12--18, 2026, Rio de Janeiro, Brazil}
\acmDOI{10.1145/3744916.3787788}
\acmISBN{979-8-4007-2025-3/2026/04}

\newcommand{\agent}{\textsc{EmbedAgent}}
\newcommand{\benchmark}{\textsc{EmbedBench}}
\newcommand{\problemNum}{\textsc{126}}

\usepackage{graphicx}
\usepackage{listings}
\usepackage{colortbl}
\usepackage{algorithm}
\usepackage{algorithmic}
\usepackage{multirow}
\usepackage{subfigure}
\usepackage{caption}
\usepackage{xcolor}

\usepackage{pifont}
\usepackage{amsfonts}
\usepackage{fontawesome}
\usepackage{enumitem}

\newfloat{listing}{h}{lop}  

\definecolor{codegreen}{rgb}{0,0.4,0}
\definecolor{codegray}{rgb}{0.5,0.5,0.5}
\definecolor{codepurple}{rgb}{0.58,0,0.82}
\definecolor{backcolour}{rgb}{0.95,0.95,0.92}
\definecolor{darkgreen}{rgb}{0,0,0}
\definecolor{lightblue}{RGB}{115, 192, 222}

\lstdefinestyle{mystyle}{   
    commentstyle=\color{codegreen},
    keywordstyle=\color{magenta},
    numberstyle=\color{codegray},
    stringstyle=\color{codepurple},
    morekeywords={seg_code},
    morecomment=[l]{;;}
}

\lstdefinestyle{acmC}{
  language=C,
  basicstyle=\ttfamily\small,
  keywordstyle=\ttfamily,
  commentstyle=\ttfamily\itshape\color{codegreen},
  stringstyle=\ttfamily,
  numbers=none,              
  frame=none,                
  breaklines=true,
  breakatwhitespace=true,
  tabsize=2,
  showstringspaces=false
}

\definecolor{bg}{HTML}{F8F9FB}  
\definecolor{bgc}{HTML}{FCF6E4}

\begin{document}

\title{\agent: Benchmarking Large Language Models \\in Embedded System Development}

\author{Ruiyang Xu}
\authornote{Both authors contributed equally to this research.}
\authornote{Also affiliated with Chinese Information Processing Laboratory, Institute of Software, Chinese Academy of Sciences, Beijing, China}
\email{xuruiyang2022@iscas.ac.cn}
\affiliation{%
  \institution{University of Chinese Academy of Sciences}
  \city{Beijing}
  \country{China}
}

\author{Jialun Cao}
\authornotemark[1]
\email{jcaoap@cse.ust.hk}
\affiliation{%
  \institution{The Hong Kong University of Science and Technology}
  \city{Hong Kong}
  \country{China}
}

\author{Mingyuan Wu}
\email{wumy@pcl.ac.cn}
\affiliation{%
  \institution{Peng Cheng Laboratory}
  \city{Shen Zhen}
  \country{China}
}

\author{Wenliang Zhong}
\authornotemark[2]
\email{zhongwenliang2024@iscas.ac.cn}
\affiliation{%
  \institution{University of Chinese Academy of Sciences}
  \city{Beijing}
  \country{China}
}

\author{Yaojie Lu}
\authornotemark[2]
\authornote{Corresponding Author}
\email{luyaojie@iscas.ac.cn}
\affiliation{%
  \institution{Institute of Software, Chinese Academy of Sciences}
  \city{Beijing}
  \country{China}
}

\author{Ben He}
\authornotemark[2]
\authornotemark[3]
\email{benhe@ucas.edu.cn}
\affiliation{%
  \institution{University of Chinese Academy of Sciences}
  \city{Beijing}
  \country{China}
}

\author{Xianpei Han}
\authornotemark[2]
\email{xianpei@iscas.ac.cn}
\affiliation{%
  \institution{Institute of Software, Chinese Academy of Sciences}
  \city{Beijing}
  \country{China}
}

\author{Shing-Chi Cheung}
\email{scc@cse.ust.hk}
\affiliation{%
  \institution{The Hong Kong University of Science and Technology}
  \city{Hong Kong}
  \country{China}
}

\author{Le Sun}
\authornotemark[2]
\email{sunle@iscas.ac.cn}
\affiliation{%
  \institution{Institute of Software, Chinese Academy of Sciences}
  \city{Beijing}
  \country{China}
}

\renewcommand{\shortauthors}{Trovato et al.}

\begin{abstract}

Large Language Models (LLMs) have shown promise in various tasks, yet few benchmarks assess their capabilities in embedded system development. 
In this paper, we introduce \agent, a paradigm designed to simulate real-world roles in embedded system development, such as \textit{Embedded System Programmer}, \textit{Architect}, and \textit{Integrator}.
This paradigm enables LLMs to be tested in tasks that bridge the gap between digital and physical systems, allowing for a more comprehensive assessment of their capabilities.
To evaluate LLMs on these tasks, we propose \benchmark, the first comprehensive benchmark for embedded system programming, circuit design, and cross-platform migration.
\benchmark~ consists of 126 cases, covering 9 electronic components across 3 hardware platforms.
Through extensive experiments on 10 mainstream LLMs, we uncover several key findings.
Surprisingly, despite the simplicity of the cases, DeepSeek-R1 achieves only a 55.6\% pass@1 rate when provided with schematic information, and 50.0\% when tasked with generating the schematics itself.
In the cross-platform migration tasks, LLMs show relatively strong performance with MicroPython on the Raspberry Pi Pico (with the top model achieving 73.8\% pass@1), but perform poorly on ESP-IDF, where the best model reaches only 29.4\% pass@1.
Interestingly, we observe that general-purpose chat LLMs like DeepSeek-V3 often fail to utilize relevant pre-trained knowledge in this domain, while reasoning LLMs tend to overthink and overlook efficient knowledge during pretraining.
Based on these insights, we propose two strategies—retrieval augmented generation and compiler feedback-to enhance LLM performance. 
These strategies result in significant improvements, with Deepseek-R1 reaching a 65.1\% pass@1 with correct schematics, and 53.1\% without. Additionally, the accuracy of the Arduino to ESP32 migration task improves from 21.4\% to 27.8\%.

\end{abstract}



\settopmatter{printfolios=true}

\maketitle

\section{Introduction}

\begin{figure}[th]
\centering
\includegraphics[width=1\linewidth]{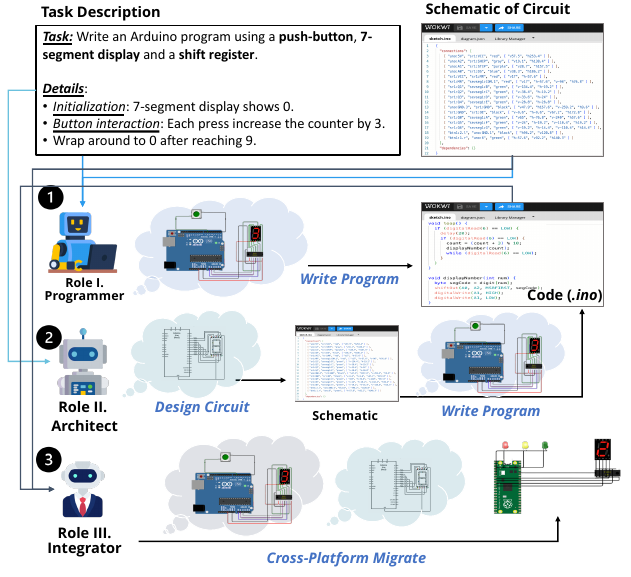}
\setlength{\abovecaptionskip}{-5pt}
\setlength{\belowcaptionskip}{-15pt}
\caption{Three \textbf{Settings} of \agent: \ding{202} \textbf{Embedded System Programmer}: Given the task description and schematic of circuit, LLMs are expected to write Arduino code. \ding{203} \textbf{Embedded System Architect}: Given the task description, LLMs are expected to design the circuit and write the code. \ding{204} \textbf{Embedded System Integrator}: Given the schematic of circuit and code of one hardware, LLMs are expected to migrate the design of circuit and code to another hardware platform.}
\label{fig:task_def}
\end{figure}

Large Language Models (LLMs) have demonstrated remarkable expertise across various software engineering tasks, such as code generation ~\citep{liu2024your,du2024evaluating}, defect detection ~\citep{yang2024large}, program repair ~\citep{xia2023keep,zhong2024ldb,hu2024leveraging} and code translation ~\citep{xu2024cruxeval}.
To assess these abilities of LLMs, benchmarks like HumanEval ~\citep{chen2021evaluating} and SWE-bench ~\citep{jimenez2024swebench} have been introduced\textcolor{darkgreen}{, along with benchmark-connected agents to solve these benchmarks ~\citep{yang2024swea,team2025kimi}}, offering valuable insights into the strengths and limitations of LLMs.

\textbf{\textit{Research gap}} -- However, as a key carrier connecting the digital LLMs to the physical world~\citep{jimenez2013introduction,wolf2010high}, 
there are few benchmarks to evaluate the capabilities of LLMs in the development of embedded systems. Prior studies~\cite{englhardt2024exploring,yang2024embedgenius} have primarily focused on code generation for embedding application, leaving their abilities in circuit design and cross-platform migration largely unexplored.

\textit{\textbf{\benchmark}} -- To address the research gap, we construct \benchmark, a new benchmark designed to evaluate the capabilities of LLMs on fundamental embedded system tasks.
\textcolor{darkgreen}{Distinct from prior benchmarks that only assess single-stage programming ability, \benchmark~introduces \textbf{\agent}, which includes multi-stage and hardware-aware evaluation settings, enabling end-to-end assessment from circuit design to cross-platform deployment.
This makes~\benchmark the first benchmark to systematically measure how well LLMs can bridge the gap between digital reasoning and physical device interaction, as illustrated in Figure~\ref{fig:task_def}:}

\begin{itemize}[leftmargin=*]
    \item[\ding{202}] \textbf{Embedded System Programmer}: Given a task description and a circuit schematic, LLMs
     are expected to write embedded system code (e.g., Arduino code).
    \item[\ding{203}] \textbf{Embedded System Architect}: \textcolor{darkgreen}{Given a task description, LLMs
     must first read and interpret the problem description, select the appropriate components, design the circuit schematics, and then write code corresponding to the designed schematics. In other words, the architect is responsible for the entire embedded system design process.}
    \item[\ding{204}] \textbf{Embedded System Integrator}: Given the code and the circuit schematic  for one hardware platform, LLMs
     are expected to migrate both the circuit design and code to another platform (Arduino, ESP32, and Raspberry Pi Pico). 
\end{itemize}
For each setting, there are manually constructed \problemNum\ cases in \benchmark, including task description, reference solution and automated correctness validation.

However, a comprehensive evaluation of LLMs' capabilities in embedded system development presents several challenges.
First, representing circuit schematics in a format that LLMs can understand and generate is challenging, since LLMs \textcolor{darkgreen}{often struggle to comprehend the functions and interactions of physical components.}
Second, verifying the correctness of the generated code is difficult.
Previous work often evaluates correctness through serial output~\cite{englhardt2024exploring,yang2024embedgenius}, but this approach can be misleading,  as LLMs may produce plausible outputs without ensuring correct hardware behavior.
Finally, verifying LLM-generated solutions at scale remains costly and inefficient, as manual assembly and physical testing are time-consuming and resource-intensive.

\textbf{\textit{Design for End-to-End Automated Evaluation}} -- To overcome the above challenges,  we design a comprehensive evaluation from \textbf{\textit{representation}}, \textbf{\textit{evaluation}}, and \textbf{\textit{pipeline}} aspects.
First, we design an \textbf{\textit{interpretable circuit representation}} that allows LLMs to understand and generate circuit structures effectively, bridging the gap between natural language inputs and hardware representation.
Second, to overcome the limitations of traditional evaluation based on serial output, we introduce a \textbf{\textit{hardware-driven evaluation approach}}, where the behavior of virtual hardware in the Wokwi \footnote{\url{https://wokwi.com/}} environment is monitored in real-time to verify the correctness of the generated code. 
Thrid, to tackle the inefficiency of manual validation, we develop an \textbf{\textit{end-to-end automated evaluation pipeline}}, which maps LLM-provided schematics to simulated connections within the virtual environment, ensuring an efficient and scalable evaluation process.

\textbf{\textit{Study}} -- Through extensive experiments on 10 mainstream LLMs, we uncover several key findings.
First, although our benchmark focuses on relatively basic tasks in embedded system development, these problems remain challenging for state-of-the-art LLMs.
For example, even when provided with correct circuit schematics, the best-performing model, Deepseek-R1, achieves only 55.6\% pass@1. When the correct schematics are not provided, its performance drops to 50\% pass@1. In the cross-platform migration task (from Arduino to ESP32), the best result is achieved by Claude 3.7 Sonnet (Thinking), with a pass@1 score of only 29.4\%.
Second, we observed that different types of LLMs exhibit different weaknesses in this domain. For example, chat LLMs often struggle to flexibly apply their pretrained knowledge.
In contrast, reasoning LLMs tend to approach problems from a very low-level perspective and often fail to effectively leverage the higher-level knowledge encoded during pretraining. Surprisingly, in some cases, a reasoning LLM that design the circuit by itself can outperform the same model when given a predefined schematic. 

Building on these insights, we explore two strategies to improve LLM performance in embedded system development tasks:
1) \textit{Retrieval augmented generation}, which enhances LLMs by incorporating previous experience;
2) \textit{Complier feedback}, which helps correct syntax errors in the generated code.
With these methods, Deepseek-R1 improves to 65.1\% pass@1 when provided with correct circuit schematics, and 53.1\% without them.
For the Arduino to ESP32 migration task, accuracy improves from 21.4\% pass@1 to 27.8\%.

The contribution of this paper includes:

\ding{52} \textbf{Benchmark \benchmark:} We introduce \benchmark, a benchmark for embedded system development. To the best of our knowledge, this is the first comprehensive benchmark designed to assess LLMs' capability in embedded system programming, circuit design, and cross-platform migration. It includes of 126 cases, covering 9 electronic components across 3 hardware platforms.

\ding{52} \textbf{\textit{Executable Context and automated validation mechanism}}: We propose an efficient and automated evaluation pipeline for assessing embedded systems developed by LLMs. This framework streamlines the testing process, ensuring consistency and reliability.

\ding{52} \textbf{\textit{Comprehensive Study and Vision}}: We conduct a comprehensive evaluation of 10 mainstream LLMs on our benchmark and uncover several insightful findings regarding their strengths and limitations. Based on these findings, we propose two effective methods to improve LLM performance in embedded system tasks.

\section{Background}

\begin{figure}[t!]
\centering
\includegraphics[width=1\linewidth]{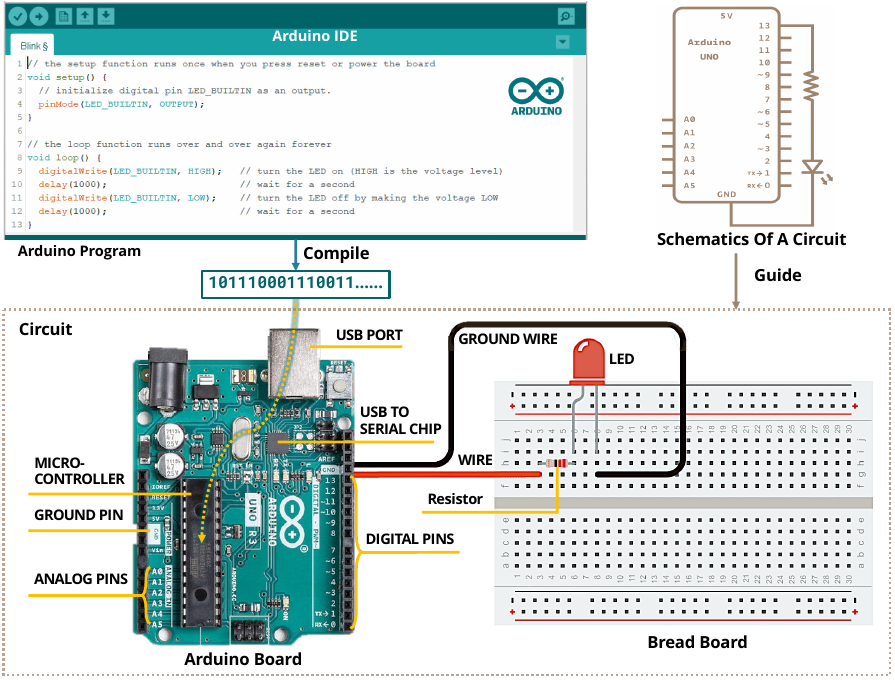}
\caption{\textbf{Arduino Workflow}. The workflow requires a combination of hardware and software. On the hardware side, an Arduino board, electronic components, and a circuit schematic are needed. On the software side, code must be written based on the schematic and then uploaded to the board via USB port and a USB-to-serial chip, finally processed in the microcontroller.}
\label{fig:arduino}
\end{figure}

In this section, we provide background information on the workflow of embedded systems and introduce Wokwi, a virtual circuit simulation platform.

\subsection{Embedded System Workflow}

Embedded systems are specialized computing systems that are based on hardware and driven by software. They are widely used in areas such as smart homes, robotics, and the Internet of Things (IoT). These systems serve as an important bridge between the virtual world and the physical world. Building a complete embedded system typically requires a combination of both hardware and software. Figure \ref{fig:arduino} illustrates the process of building a simple embedded system using an Arduino development board and an LED.

On the hardware side, it is necessary to prepare the development board and the required components. In this case, the components are an Arduino board and an LED. Then, based on the components and the serial port information of the board, the circuit schematic is constructed.

On the software side, code is written according to the schematic. In the figure, the Arduino IDE is used to write code that makes the LED blink every second. This code is compiled into a binary file and uploaded to the development board, enabling interaction between the virtual world and the physical world.

\subsection{Virtual Circuit Simulation Platform Wokwi}
Wokwi\cite{wokwi} is an online simulation platform for electronic circuits and embedded systems. It is primarily used to simulate the behavior of various microcontrollers such as Arduino, ESP32, and Raspberry Pi Pico, along with peripheral electronic components like sensors, LEDs, and displays. It provides a convenient virtual environment that allows developers to quickly build, test, and debug projects directly in the browser, enabling code and circuit design verification without the need for physical hardware.

\begin{figure}[t!]
\centering
\includegraphics[width=1\linewidth]{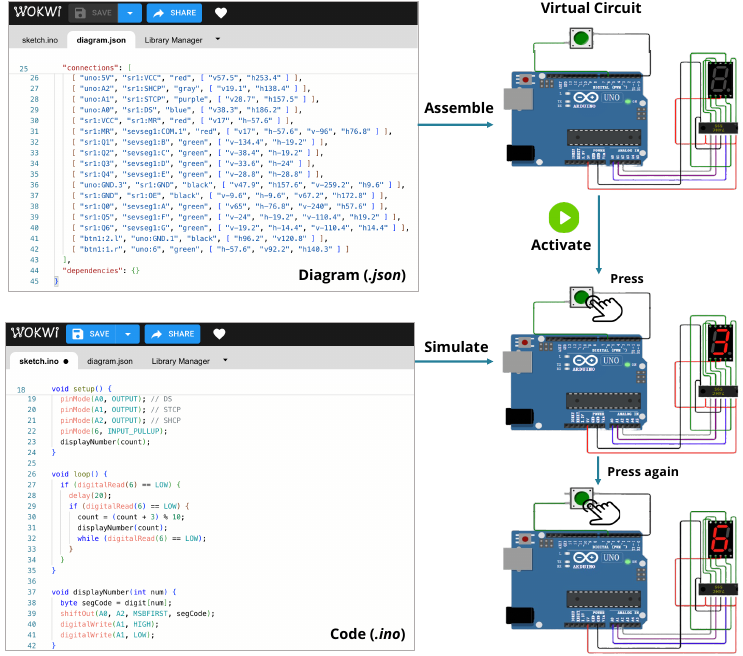}
\caption{\textbf{Illustration of Wokwi~\cite{wokwi} - A Virtual Circuit Simulation}. The upper part (i.e., Diagram) shows how the virtual circuit is denoted in Wokwi, the lower part (i.e., Code) shows how the Arduino code simulates the virtual circuit. Once the virtual circuit is activated, it will be virually powered. According to the Arduino code, the number shows in 7-segment display increase once the button has been pressed. }
\label{fig:wokwi}
\end{figure}

Figure \ref{fig:wokwi} shows an example of developing an embedded system using the Wokwi platform. Similar to real-world development, creating an embedded system in Wokwi also involves both software and hardware. On the hardware side, appropriate development boards and necessary components are selected, followed by schematics. The schematics are represented in the Diagram (.json) shown in Figure \ref{fig:wokwi}, where a structured file records the connections.

On the software side, the corresponding Arduino code is written, as shown in the Code (.ino) section of the figure. Once the code is complete, clicking the ``Activate'' button on the website compiles the program. Users can interact with the embedded system through their computer to determine whether the expected functionality has been achieved.

\section{Benchmark Construction}

In this section, we provide a detailed introduction to the construction process of our benchmark. It can be divided into three main steps. First, determining the components required for each problem. Second, formulating appropriate problems based on the hardware usage. Finally, constructing test cases according to the problems. We also present statistics of our benchmark and analyze its quality.

\subsection{Electronic Component Selection}
To ensure balanced utilization of components across problems, we selected seven primary hardware types as seed components: LED, RGB LED, Push Button, 7-Segment Display, LED Bar Graph, Slide Potentiometer and Servo. Due to space limitation, the detailed descriptions of each component are provided in Appendix A
in \cite{artifact}. As illustrated in Figure \ref{fig:pipline} Step I, each problem is assigned a distinct set of electronic components. For each problem, we randomly select 1 to 3 components from the seven primary types. Each selected component is then randomly assigned a usage count between 1 and 3. We sample each unique hardware combination twice, resulting in a total of 126 cases, each with a unique distribution of components. An example problem is shown in the Task Description of Figure \ref{fig:pipline} step II, where the sampled configuration includes two types of components (7-Segment Display and Push Button), each hardware use one instance.

\subsection{Problem Construction}
After determining the required electronic components for each problem, we formulate the problem descriptions through human annotation based on these components. Each problem must satisfy two key conditions. First, all selected electronic components must be actively utilized within the problem. Second, any given system state must be maintained for at least 2 seconds, \textcolor{darkgreen}{which corresponds to the minimum interval that ensures stable and accurate results, as verified by repeated 10-time testing on sampled data.}

Each problem description is composed of two parts. The first part provides a high-level overview, specifying the electronic components involved and the task's general objective. The second part outlines the specific operations and the expected states of each electronic component following those operations. \textcolor{darkgreen}{Each problem description is manually reviewed to ensure that every task is meaningful and coherent.}

\begin{figure*}[t!]
\centering
\includegraphics[width=0.99\textwidth]{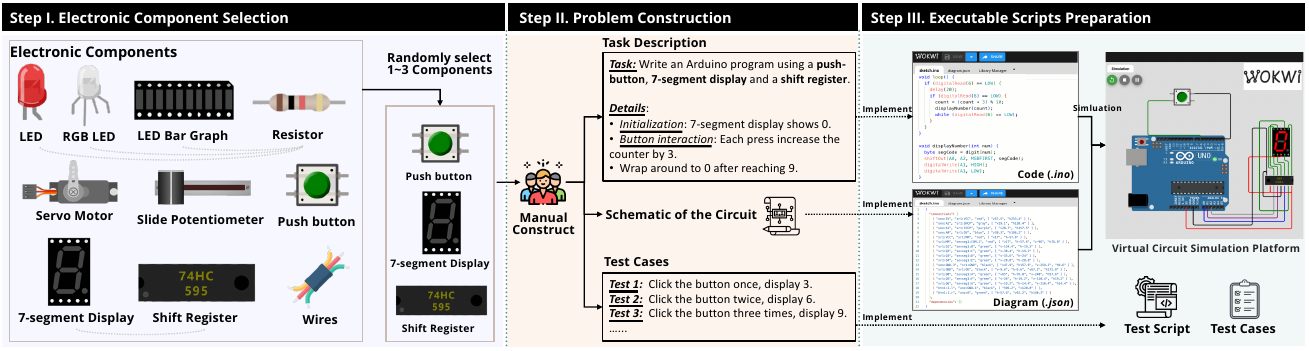}
\caption{Data Preparation Pipeline}
\label{fig:pipline}
\end{figure*}

A simplified version is shown in the Task Description of Figure \ref{fig:pipline}, Step II. The problem description begins by stating that the electronic components include a button and a 7-segment display. The task is to implement a counter. It then specifies the detailed behavior: The initial state of the 7-segment display should show the number 0, and each button press should increment the counter by 3.

Based on the formulated problems, we develop corresponding test cases, with each problem containing 3 to 5 test cases. These test cases are designed to verify whether the code generated by the LLMs satisfies all the requirements specified in the problem description. Additionally, we provide correct reference implementations to ensure that each problem is solvable.

For example, in Figure \ref{fig:pipline}, Step II, to verify the correctness of the code, we need to check whether the output for the $n$th button click is equal to $(n*3) mod 10$. Accordingly, the first three test cases check the behavior for 1 to 3 button clicks. A fourth test case simulates 20 button presses to verify whether the code handles extended sequences and edge conditions correctly, ensuring comprehensive evaluation.

\subsection{Executable Scripts Preparation}
Unlike previous work that relies on physical hardware and determines correctness through serial output, we use Wokwi simulation platform to observe the real-time state of each electronic component at any given moment. This capability is essential for evaluating embedded systems, as the primary goal of embedded systems is to ensure that the electronic components perform the expected operations.

We use an automated submission bot to submit the code generated by LLMs and monitor the electronic components to ensure the behavior aligns with the specified requirements. Each problem is associated with a unique script containing all the test cases defined in Step II, which are used to verify correctness. For instance, in Step III of Figure \ref{fig:pipline}, the validation process involves checking whether the 7-segment display shows the expected number after the push button has been pressed a specified number of times. \textcolor{darkgreen}{Leveraging the executable scripts and simulation platform, we enable automatic verification for each problem. In addition, every problem is accompanied by a golden solution that passes all verification tests, ensuring that each problem is solvable.}

\subsection{Quality Analysis}

Each problem, along with its corresponding solutions and test cases, is created by two volunteers. After the initial development, another two volunteers are responsible for reviewing the content to ensure its quality and accuracy.\textcolor{darkgreen}{~Each volunteer holds a bachelor's degree in computer science and has 1–3 years of experience in embedded programming, ensuring domain expertise and practical relevance in task design.} Annotators should follow these main guidelines: (1) The problem description must fully utilize all the specified electronic components. (2) The provided test cases must be sufficient to verify the correctness of the problem, ensuring that the reference code passes all of them. (3) Any states that require verification must be maintained for at least two seconds to avoid the impact of potential network delays. This multi-stage review process helps ensure the correctness, reliability, and overall quality of the questions. Further details on annotation rules are provided in Appendix C of \cite{artifact}. Appendix D presents detailed statistics of our benchmark, and Appendix F discusses its suitability as a benchmark.

\section{Experiments}

\subsection{Experiment Setup}
\subsubsection{Evaluation Tasks}
To systematically evaluate embedded systems, we propose three task settings designed to assess LLMs' embedded system development capabilities under different conditions: Settings \ding{202},\ding{203} and \ding{204} which is shown in Figure \ref{fig:task_def}. \textcolor{darkgreen}{Setting \ding{204} further consists of two subtasks, namely Arduino-to-ESP32 and Arduino-to-Raspberry Pi Pico. } 

\textcolor{darkgreen}{We select Arduino Uno, Raspberry Pi Pico, and ESP32 as representative targets because they encompass 
(1) distinct hardware architectures (AVR, ARM Cortex-M, and Tensilica LX6), 
(2) diverse programming ecosystems (C++, MicroPython, and ESP-IDF), and 
(3) different abstraction levels ranging from pin-level to RTOS-level control. 
This combination ensures that the benchmark covers a wide spectrum of real-world embedded development scenarios and provides strong generalization across heterogeneous hardware and software environments.}

\begin{table}[!ht]
    \centering
    \includegraphics[width=\linewidth]{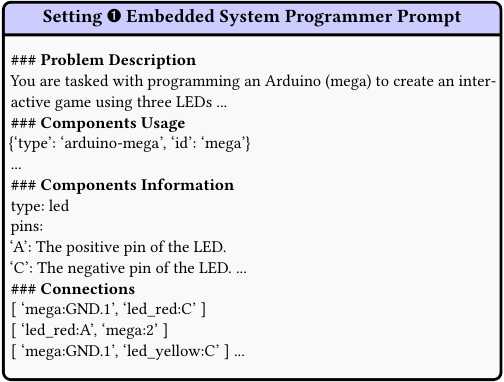}
    \caption{The prompt of Setting \ding{202} task}
    \label{tab:prompt1}
\end{table}

    




% ...





% ...





\noindent{\textbf{Setting \ding{202}}} As illustrated in Figure~\ref{fig:task_def}, this task requires LLMs to generate embedded system code based on a given problem description and a schematic configuration. Successfully completing the task demands a comprehensive understanding of the problem requirements, the functionality of each hardware component, and the roles of individual pins. This task evaluates the LLM's capability as a programmer. The process can be formalized as:
\begin{equation}
\begin{aligned}
r_i = I(P(c_{i}\ |q_i, w_i, h_i,L, B; M);u_i)
\end{aligned}
\end{equation}

\textcolor{darkgreen}{where $P(\cdot)$ denotes the probability distribution of LLM $M$ when generating code given the problem description $q_i$, schematic configuration $w_i$, hardware specifications $h_i$, programming language $L$ and the development board $B$. The function $I(\cdot)$ executes the predicted code $c_i$ on the corresponding test cases $u_i$ and returns a binary vector $r_i$, where each element indicates whether the corresponding test case passes (1) or fails (0).}

The prompt for the ``Setting \ding{202}'' tasks is shown in Tabel \ref{tab:prompt1}. First, the LLM is provided with a problem description. Next, a list of components is given, where each component is represented as a dictionary containing two keys: the component’s type and its unique identifier. Following this, detailed specifications for each component type are presented, including the available pins and their respective functions. Lastly, the schematics are described. Each connection is specified as a pair of strings in the format ``id:pin'', denoting the connection between specific pins of two components. For example, \texttt{[``uno:GND.1'', ``led1:COM'']} represents a connection between pin \texttt{GND.1} of the \texttt{uno} component and the \texttt{COM} pin of the \texttt{led1} component. The LLM is expected to generate Arduino code that utilizes the given components and schematics configuration to solve the specified problem.

\noindent\textbf{Setting \ding{203}} As shown in Figure~\ref{fig:task_def}, in this task, in addition to generating code for the embedded system, LLMs are also required to design circuits. This requires a deeper understanding of hardware components, familiarity with the functionalities of various serial ports, and a certain level of logical reasoning and spatial capabilities. This task evaluates the LLM's capability as an expert in embedded system development. The process can be formulated as:
\begin{equation}
\begin{aligned}
r_i = I(P(w_i, c_{i}\ |q_i, h_i,L, B; M);u_i)
\end{aligned}
\end{equation}

\textcolor{darkgreen}{Here, the LLM must jointly generate both the schematic $wi$ and the corresponding code $ci$ based on the problem description $q_i$ and hardware information $h_i$. The correctness is verified by executing $c_i$ on the generated schematic using the test cases $u_i$ through the validation function $I(\cdot)$.}

As shown in Table \ref{tab:prompt2}, the prompt is constructed similarly to the ``Setting \ding{202}'' task. LLMs are given a task description without schematics of circuits. 
To format the output, we include example connections to guide the LLM in constructing an appropriate schematic in the required format. The LLM is expected to return both the schematic for the given components and the code to control the components and complete the task.

\begin{table}[t]
    \centering
    \includegraphics[width=\linewidth]{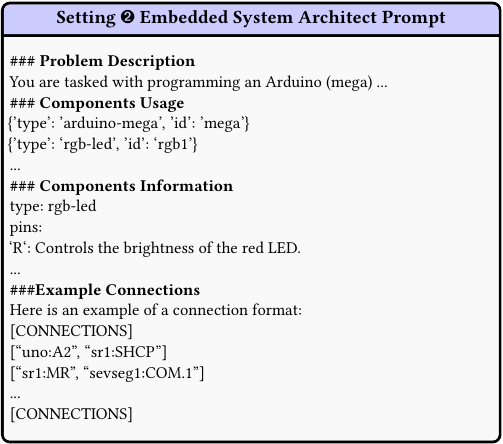}
    \caption{The prompt of Setting \ding{203} task}
    \label{tab:prompt2}
\end{table}

    





% ...





% ...







% ...




\noindent{\textbf{Setting \ding{204}}} As shown in Figure~\ref{fig:task_def}, the ``\textit{Setting \ding{204}}'' task requires LLMs to migrate both the schematics and code from a given source platform to a target platform. The LLM must consider the similarities and differences between platforms, including serial ports, programming languages, and the syntax of various function calls. This process can be formalized as:
\begin{equation}
\begin{aligned}
r_i = I(P(w_i', c_{i}'\ |h_i',L', B',w_i,c_i; M);u_i)
\end{aligned}
\end{equation}

\textcolor{darkgreen}{In this setting, the LLM is given an existing schematic–code pair $(w_i, c_i)$ from a source platform and must generate corresponding outputs $(w_i', c_i')$ for a target platform with hardware $h_i'$, language $L'$, and board $B'$. The validation function $I(\cdot)$ checks whether the migrated code passes all functional test cases on the new platform.} Detailed information of each platform used in this task is provided in Appendix B
in \cite{artifact}. 

Unlike the ``Setting \ding{202}'' and ``Setting \ding{203}'' tasks. As shown in Table \ref{tab:prompt3}, the ``Setting \ding{204}'' task does not include an explicit problem description. Instead, because it involves platform migration, the input includes the Arduino code and schematics connections. The component usage and specifications remain consistent with the previous two tasks, with the addition of the target platform's details. The LLM must first translate the schematics from Arduino to the target platform and then generate equivalent code that replicates the same functionality.

\begin{table}[t]
    \centering
    \includegraphics[width=\linewidth]{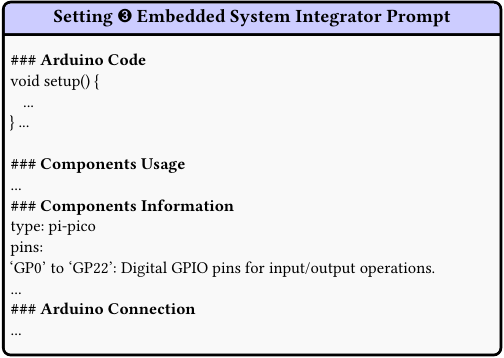}
    \caption{The prompt of Setting \ding{204} task}
    \label{tab:prompt3}
\end{table}
\subsubsection{Evaluation Method}

    





% ...





% ...


% ...


We use pass@1 as the evaluation metric following prior work~\cite{kulal2019spoc,chen2021evaluating}. \textcolor{darkgreen}{This metric measures the proportion of problems for which the LLM’s first generated solution is correct. A problem is considered successfully solved if the LLM passes all test cases. We reasoned that the ultimate goal of embedded system design is for the system to fully meet user requirements rather than only a subset of them. Moreover, in real-world scenarios, users typically query an LLM once for a given problem instead of generating multiple attempts. Therefore, we selected pass@1 as the final evaluation metric. To ensure reproducibility and stability, we adopted greedy decoding with the temperature set to 0, which ensures that the LLM produces the same output each time.}

\subsubsection{LLMs for evaluation}
We select 10 state-of-the-art LLMs spanning 6 different LLM families, and categorize them into three types for evaluation: \textbf{\textit{Reasoning LLMs}} (O3-mini\cite{o3_mini}, Claude 3.7 Sonnet (Thinking)\cite{claude}, DeepSeek-R1\cite{guo2025deepseek}, QWQ\cite{qwq32b}), \textbf{\textit{SFT-Distilled Reasoning LLMs}} (DeepSeek-R1-Distill-Llama\cite{guo2025deepseek}, DeepSeek-R1-Distill-Qwen\cite{guo2025deepseek}), and \textbf{\textit{Chat LLMs}} (DeepSeek-V3\cite{deepseekai2024deepseekv3technicalreport}, Llama-3.3-Instruct\cite{meta_llama_3}, Starcoder2\cite{lozhkov2024starcoder}, Qwen2.5-Instruct\cite{qwen2.5}).

\subsection{Overall Result}
Table \ref{tab:overallResult} presents the pass@1 evaluation results for different types of LLMs, arranged in descending order based on their parameter size. The key observations are as follows: 

\textbf{\textit{1) \benchmark~ is challenging for studied LLMs.}} In both the ``Setting \ding{202}'' and ``Setting \ding{203}'' tasks, \textcolor{darkgreen}{although our benchmark problems are relatively simple, similar to HumanEval, LLMs still achieve only around 50\% accuracy on average}, the highest-performing LLM, DeepSeek-R1, achieves only 55.6 and 50.0 pass@1, respectively. For the ``cross-platform'' task, most LLMs struggle to transfer Arduino code to ESP32's ESP-IDF effectively. Even the best-performing LLM, Claude 3.7 Sonnet (thinking), reaches only 29.4 pass@1. Transferring code to the Raspberry Pi Pico using MicroPython is slightly more successful, with a pass@1 of 73.8 at best.

\textbf{\textit{2) Reasoning LLMs outperform Chat LLMs. However, SFT-distilled reasoning LLMs show mixed results compared to their base versions.}} For example, QwQ surpasses Qwen2.5-Instruct, across all tasks. However, DeepSeek-R1-Distill-Qwen does not show a significant improvement over its base LLM. Additionally, DeepSeek-R1 consistently outperforms DeepSeek-V3 in all tasks.

\begin{table}[t!]
\renewcommand\arraystretch{1.2}
  \resizebox{1\linewidth}{!}{
  \centering
    \begin{tabular}{l|c|cccc}
    \toprule
    \multicolumn{1}{c|}{\multirow{4}[6]{*}{Models}} & \multirow{4}[6]{*}{Size} & \multicolumn{4}{c}{Task} \\
\cmidrule{3-6}          &       & \multicolumn{1}{c|}{Setting \ding{202}} & \multicolumn{1}{c|}{Setting \ding{203}} & \multicolumn{2}{c}{Setting \ding{204}
} \\
\cmidrule{3-6}          &       & \multicolumn{1}{c|}{Arduino} & \multicolumn{1}{c|}{Arduino} & \multicolumn{1}{c|}{ESP32} & Pi Pico \\
          &       & \multicolumn{1}{c|}{C++} & \multicolumn{1}{c|}{C++} & \multicolumn{1}{c|}{ESP-IDF} & MicroPy \\
    \midrule
    \multicolumn{6}{c}{\textit{Reasoning LLM}} \\
    \midrule
    Claude 3.7 & \textbackslash{} & 48.4 (+4,8)  & 48.4  & \textbf{ 29.4 } & \textbf{73.8} \\
    O3-mini & \textbackslash{} & 47.6  & 38.9  & 16.7  & 61.9  \\
    DS-R1 & 671B/31B & \textbf{ 55.6 } & \textbf{ 50.0 } & 21.4  & 60.3  \\
    QWQ   & 32B   & 39.7  & 50.0  & 4.0   & 52.4  \\
    \midrule
    \multicolumn{6}{c}{\textit{SFT-Distilled Reasoning LLM}} \\
    \midrule
    DS-R1-Dist. & 70B   & 35.7  & 23.8  & 0.0   & 11.1  \\
    DS-R1-Dist. & 32B   & 27.0  & 11.1  & 3.2   & 27.8  \\
    \midrule
    \multicolumn{6}{c}{\textit{Chat LLM}} \\
    \midrule
    DS-V3 & 671B/31B & 37.3  & 25.4  & 2.4   & 46.8  \\
    Llama-3.3-Ins. & 70B   & 32.5  & 21.4  & 0.0   & 23.0  \\
    Qwen2.5-Ins. & 32B   & 25.4  & 23.8  & 3.2   & 33.3  \\
    Starcoder2 & 15B   & 11.1  & 2.4   & 0.0   & 4.0  \\
    \bottomrule
    \end{tabular}%
    }
  \caption{pass@1(\%) results of each LLMs. Abbreviations: sch (schematics), DS(DeepSeek), Claude 3.7(Claude 3.7 Sonnet (thinking)), DS-R1-Dist. 70B (DeepSeek-R1-Distill-Llama), DS-R1-Dist. 32B (DeepSeek-R1-Distill-Qwen), Ins. (Instruct). Ds-R1 and DS-V3 are MOE LLMs with 32B parameters activated during inference.}
 \label{tab:overallResult}%
\end{table}%

\section{Analysis}
We first analyze why LLMs struggle to construct embedded systems by investigating concrete cases. \textcolor{darkgreen}{Next, we unpack the chain of thought generated by different reasoning LLMs when tackling embedded system tasks, which is essential for understanding their reasoning limitations. Based on insights from this analysis, we then propose techniques to improve LLMs’ capability in embedded system development.}

\textcolor{darkgreen}{To ensure the rigor of our qualitative observations, we conducted a manual inspection of LLMs' outputs. 
Two annotators with embedded programming experience independently reviewed the generated code and reasoning traces, 
identified reasoning patterns and failure types, and discussed discrepancies to reach consensus. }
\subsection{Case Study}
\begin{figure}[h]
  \centering
  \includegraphics[width=\linewidth]{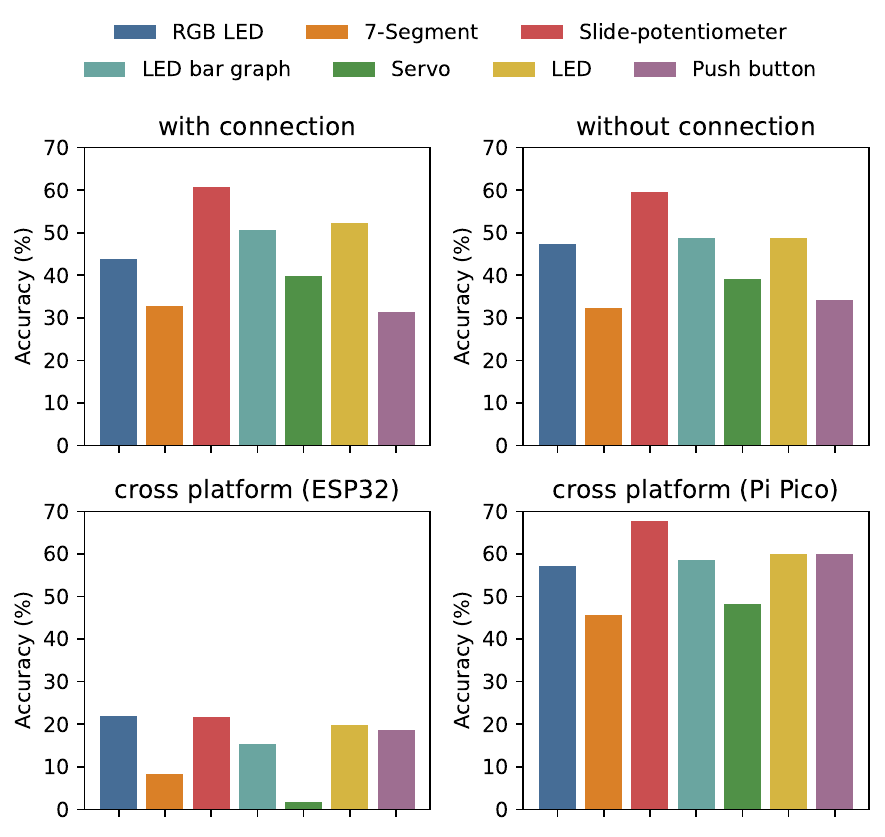}
  \caption{The average accuracy of reasoning LLMs (QwQ, DeepSeek-R1, O3-mini, Claude 3.7 Sonnet (Thinking)) on problems involving specific electronic components.}
  \label{fig:hardware_accuracy}
\end{figure}

To investigate why LLMs struggle with constructing embedded systems, we categorize problems based on the specific electronic components they involve. For instance, if a problem requires the use of both a ``Servo'' and an ``LED'', we classify it under both ``Servo'' and ``LED'' categories. We then measure the average accuracy of various reasoning LLMs (QwQ, DeepSeek-R1, O3-mini, Claude 3.7 Sonnet (Thinking)) across seven major electronic components in four different tasks. The results are presented in Figure \ref{fig:hardware_accuracy}. From the figure, we observe that LLMs struggle particularly with problems involving ``7-Segment displays'' and ``Push Buttons'' in both the ``Setting \ding{202}'' and ``Setting \ding{203}'' tasks. In the ``Setting \ding{204} (Pi Pico)'' task, LLMs generally perform well. However, in the ``Setting \ding{204} (ESP32)'' task, accuracy is consistently low across all electronic components. In the following content, we provide a detailed analysis of these challenges, focusing on the ``7-Segment'', ``Push Button'' and the ESP32 platform.

\noindent{\textbf{7-Segment Display}} \textit{LLMs struggle to determine the correct voltage levels for each port when displaying numbers (0 to 9) or characters such as ``A'' and ``P''.}

A significant portion of the errors in the ``Setting \ding{202}'' task related to 7-segment displays comes from this issue, accounting for 48 out of 116 errors (approximately 41.4\%). Similarly, in the ``Setting \ding{203}'' task, 45 out of 103 errors (approximately 43.7\%) fall into the same category.
\begin{listing}[h]
\caption{Subject-101 (QWQ)}
\label{lst:segmentExample}
\begin{lstlisting}[style=acmC]
// Lookup table for 7-segment display
// (common anode)
byte seg_code[10] = {
  0x90,0xF9,0xA2,0xB0,0x99, // 0-4
  0x92,0x82,0xF8,0x80,0x90  // 5-9
};
\end{lstlisting}
\end{listing}

An example is shown in Listing~\ref{lst:segmentExample}. The LLM uses hexadecimal values to represent the voltage levels of each segment in the 7-segment display. For instance, to represent the number 1, it uses the value \texttt{0xF9}, which corresponds to the binary pattern \texttt{0b11111001}. This means that two segments are set to low voltage (i.e., they are activated in a common-anode display), correctly displaying the digit \texttt{1}. However, the LLM struggles to consistently generate correct values for each digit. In this example, both the digits 0 and 9 are represented using the same value \textit{0x90}, which indicates an error, as they should have different segment configurations.

\noindent{\textbf{Push Button}} \textit{For push buttons, LLMs often struggle with handling button debounce.}

In the ``Setting \ding{202}'' task, nearly half of the errors related to push buttons, 43 out of 108 (approximately 39.8\%), are caused by improper debounce handling. Similarly, in the ``Setting \ding{203}'' task, 51 out of 106 errors (approximately 48.1\%) result from the same issue.
\begin{listing}[h]%
\caption{Subject-76 (Claude 3.7 sonnet (thinking))}%
\label{lst:buttonExample}%
\begin{lstlisting}[style=acmC]
// debounceDelay = 150ms
if (digitalRead(btn1Pin) == LOW && (currentTime - lastBtn1Press) > debounceDelay) {
    lastBtn1Press = currentTime;
    incrementNumber(1);
  }
\end{lstlisting}
\end{listing}

As shown in Listing~\ref{lst:buttonExample}, the LLM implements a basic debounce mechanism by checking whether the button is pressed (LOW) and ensuring that at least 150 ms have passed since the last press. However, even though the button press duration is intended to be 150 ms, various factors can make it difficult to maintain this timing precisely. As a result, this approach may lead to false triggers, where a single press is misinterpreted as multiple presses. A more reliable method would involve tracking the previous button state and updating the debounce timer only when a state change occurs (for example, from HIGH to LOW). This helps reduce the likelihood of erroneous detections.





\noindent{\textbf{ESP32}} \textit{The primary issue observed in ESP32 code generated by LLMs is the high frequency of syntax errors.}

A substantial portion of the generated code, 211 out of 504 cases (approximately 41.9\%), contains such errors. Common problems include missing or incorrect usage of relevant header files, as well as the use of functions that are incompatible with the specified version of the ESP-IDF, even when the version is clearly provided in the prompt. When comparing Python and ESP-IDF code generation, it becomes evident that current LLMs lack sufficient training in domain-specific embedded programming languages such as ESP-IDF.

\begin{center}
  \includegraphics[width=\linewidth]{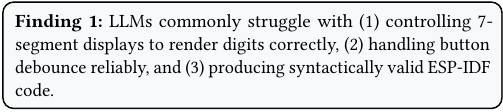}
\end{center}

\subsection{Deep thinking for Embedded System}
After conducting case studies on hardwares and platforms where LLMs are prone to make errors, we further investigate the underlying causes of these errors. Specifically, we analyze the reasoning chains produced by the reasoning LLMs to better understand how and why such errors occur in the development of embedded system.
\subsubsection{The thinking Format of each LLMs}
Given that current LLMs are trained on limited data related to embedded system development, we begin by analyzing the differences in the chains of thought produced by each reasoning LLM.

{\textbf{Claude 3.7 Sonnet (Thinking)}} \textit{devotes a significant portion of its reasoning process to debugging. This frequent and thorough debugging may help explain why Claude 3.7 Sonnet (Thinking) achieves the best performance on Setting \ding{204} tasks (e.g., ESP32), where LLMs are particularly prone to syntax errors.}
The reasoning process of Claude 3.7 Sonnet (Thinking) begins with a brief preparation phase that involves understanding the question, analyzing schematics diagrams, and outlining a plan for code generation. Following this initial stage, the LLM spends an average of 4,440 tokens (out of a total 5,703, or approximately 77.8\% of its reasoning chain) identifying and resolving issues in the generated code. As illustrated in Table~\ref{tab:claude_example}, the LLM first uses 955 tokens to generate an initial answer, then proceeds to debug the code multiple times

\begin{table}[!ht]
    \centering
    \includegraphics[width=\linewidth]{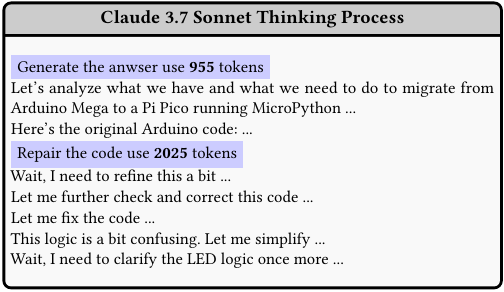}
    \caption{An example of Claude 3.7 Sonnet Thinking Process}
    \label{tab:claude_example}
\end{table}



{\textbf{DeepSeek-R1 and QWQ}}. \textit{exhibit similar reasoning patterns, marked by frequent reflection throughout each stage of the problem-solving process. A typical indicator of this reflective behavior is the word ``wait''\cite{guo2025deepseek,muennighoff2025s1}, which often signals moments of uncertainty or perceived inconsistency. This tendency toward continual self-evaluation may contribute to their strong performance on both code generation tasks
which require advanced reasoning skills.}

Compared to LLMs like Claude 3.7 Sonnet, DeepSeek-R1, and QWQ demonstrate a more introspective approach. From understanding the initial problem to analyzing potential flaws in the generated code, these LLMs repeatedly assess and revise their reasoning. The keyword ``wait'' frequently appears in their outputs as a marker of this process. On average, DeepSeek-R1 uses ``wait'' approximately 20.4 times per problem, while QWQ uses it about 40.7 times. An example is shown in \ref{tab:r1-example}. From the example, we can see that Deepseek-R1 reflects on all stages of problem understanding, code structure construction, and code testing in a balanced manner.

\begin{table}[!ht]
  \centering
  \includegraphics[width=\linewidth]{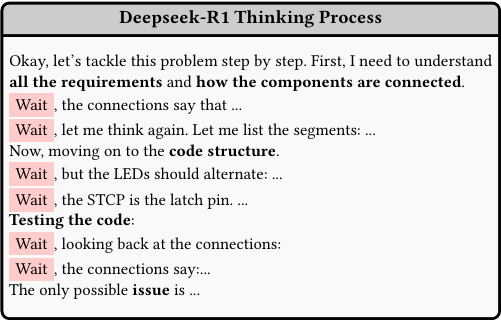}
  \caption{An example of Deepseek-R1 Thinking Process}
  \label{tab:r1-example}
\end{table}

\noindent{\textbf{SFT-distilled reasoning LLMs}} \textit{show significant instability when tackling embedded system development tasks. This instability suggest that SFT-distilled reasoning LLMs struggle with unfamiliar tasks, and may explain that in some cases these LLMs underperform compared to their base LLMs.}

Compared to their teacher LLMs, both DeepSeek-R1-Distill-Llama and DeepSeek-R1-Distill-Qwen exhibit substantially higher variability in token usage during reasoning. 
For the ``Setting \ding{202}'' and ``Setting \ding{203}'' tasks, DeepSeek-R1-Distill-Llama\textcolor{darkgreen}{ ~yields mean token counts of 5,526 (SD = 6,368) and 5,082 (SD = 5,976), respectively, while DeepSeek-R1-Distill-Qwen shows 8,815 (SD = 10,894) and 7,009 (SD = 9,450). 
In contrast, the teacher LLM averages 6,354 (SD = 3,382) and 5,951 (SD = 2,784), indicating that the distilled LLMs’ reasoning lengths fluctuate far more widely than those of their teacher. 
Here, SD denotes the standard deviation across different reasoning instances.}

On Setting \ding{204} tasks (ESP32 and Pi Pico), DeepSeek-R1-Distill-Llama averages only 696 and 569 thinking tokens, and DeepSeek-R1-Distill-Qwen averages 540 and 849, compared to DeepSeek-R1’s 3,760 and 3,848. These findings reinforce the idea that SFT-distilled reasoning LLMs are more brittle and less consistent when faced with unfamiliar problem domains.

\begin{center}
  \includegraphics[width=\linewidth]{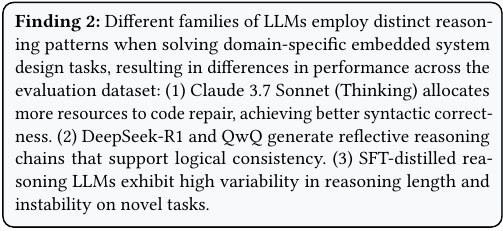}
\end{center}


\subsubsection{Why struggle to display 7-Segment?} To further explore why LLMs encounter difficulties in embedded system development, we analyze their performance on a specific hardware task: controlling a 7-segment display. This task is widely used in embedded systems and is likely represented in the training data of many LLMs. However, it also requires a multi-step reasoning process that includes hardware comprehension, pin mapping, and display logic. Based on the case study, we select three LLMs with varying capabilities, ranging from weak to strong, and evaluate how each performs when solving problems related to 7-segment displays:


\noindent \ding{192} {\textbf{Starcoder2}} is a code LLM trained on Stack v2 (67.5TB of GitHub code), which is publicly available. Using the provided tool~\cite{urlDataPortraits},
we verified that the digit table shown in Listing~\ref{lst:segmentExample} is indeed present in the Stack v2 training set. Despite this, evaluation results reveal that StarCoder2 achieved 0\% accuracy (0/45) on 7-segment display tasks, even in ``Setting \ding{202}'' task where the schematics are included. Moreover, only 22.2\% (10/45) of its generated solutions incorporated a digit table similar to the one in Listing~\ref{lst:segmentExample}. These findings indicate that \textbf{\textit{LLMs with small parameter sizes struggle to retrieve and apply pretrained knowledge effectively}} without sufficient exposure to task-specific training.

\noindent \ding{193} {\textbf{DeepSeek-V3}} is a state-of-the-art chat LLM that performs well on a variety of general tasks. However, in the ``Setting \ding{202}'' task, which uses the 7-segment display, it achieves only 6.7\% accuracy (3/45), despite including a digit table in 93.3\% (42/45) of its outputs. While DeepSeek-V3 successfully activates knowledge related to 7-segment display data present in its pretraining corpus, only 15.5\% (7/45) of its generated solutions correctly adapted the digit table to a common anode configuration as required by the task. Even fewer solutions incorporated more complex changes in response to the problem's requirements. These results indicate that while larger LLMs trained on more extensive datasets \textbf{\textit{demonstrate an increased ability to recall relevant pretraining knowledge}}, they still face challenges in flexibly applying that knowledge based on the task-specific contexts.

\noindent \ding{194} {\textbf{DeepSeek-R1}} is a reasoning LLM that differs from chat LLMs by employing \textit{a longer chain-of-thought} before output. This reasoning ability enables it to outperform the previous LLMs, achieving 44.4\% accuracy (20/45) on the ``Setting \ding{202}'' task. However, reasoning-based LLMs also face challenges in 7-segment tasks. Unlike its base LLM, DeepSeek-V3, which only reuses pretraining knowledge with minimal adaptation, DeepSeek-R1 tends to over-analyze. On average, it uses approximately \texttt{4429} tokens ((4429/10342) \texttt{42.8}\% of the total chain of thought) focusing on binary representations of digits—an unnecessary detour for this relatively simple task. This excessive reasoning often leads to errors, with some chains of thought even stating, ``But this is tedious.'' (See Appendix G
in \cite{artifact}). A more effective approach would be to retrieve a known digit table (e.g., Listing~\ref{lst:segmentExample}), and then adapt it based on the task requirements. These results show that reasoning LLMs may \textbf{\textit{entangle in unnecessarily complex thought processes}}, limiting their ability to apply pretrained knowledge efficiently in domain-specific tasks.


\begin{center}
  \includegraphics[width=\linewidth]{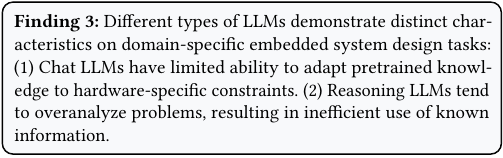}
\end{center}

\subsubsection{The gap between Setting \ding{202} and Setting \ding{203}} From the results in Table \ref{tab:overallResult}, we observe an interesting phenomenon. For chat LLMs and SFT-distilled reasoning LLMs, performance on the ``Setting \ding{202}'' task is significantly better than on the ``Setting \ding{203}'' task. However, for reasoning LLMs, the ``Setting \ding{202}'' task does not show a clear advantage over the ``Setting \ding{203}'' task. In fact, for the QWQ, the ``Setting \ding{203}'' performance is even better than the ``Setting \ding{202}'' result. In this section, we aim to analyze why the performance gap between these two tasks varies across different types of LLMs.

\begin{table}[!ht]
    \centering
    \includegraphics[width=\linewidth]{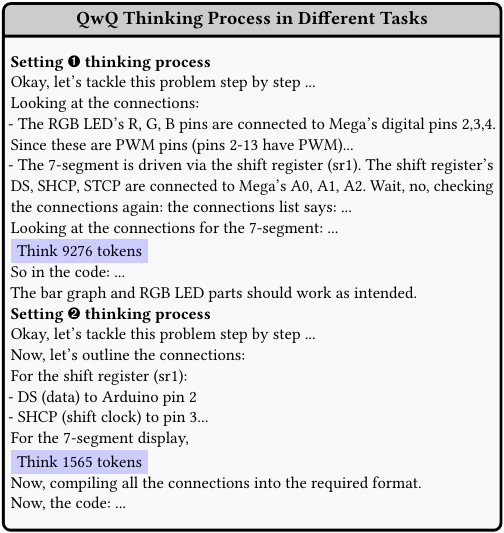}
    \caption{An example of QWQ Thinking Process}
    \label{tab:qwq-example}
\end{table}

As discussed in Section 4.2.1, reasoning LLMs such as DeepSeek-R1 and QwQ often use the keyword ``wait'' to indicate moments of uncertainty or inconsistencies in their reasoning process. Therefore, the frequency of the word ``wait'' can serve as a rough proxy for the LLM's perceived difficulty in handling a task. Interestingly, DeepSeek-R1 used the word ``wait'' 3,369 times in the ``Setting \ding{203}'' task, but this number increased to 4,503 in the ``Setting \ding{202}'' task. The discrepancy is even more pronounced for QwQ, which used ``wait'' 5,779 times in the ``Setting \ding{203}'' setting and 9,782 times in the ``Setting \ding{202}'' setting. Contrary to expectations, these results suggest that for LLMs with strong reasoning capabilities, it may actually be more effective to allow them to perform the schematics themselves, rather than supplying the correct schematics in advance.

To better understand this phenomenon, we examined the cases that QWQ successfully solved in the ``Setting \ding{203}'' task but failed to solve in the ``Setting \ding{202}'' setting. We found that the LLM often discovered alternative schematic schemes that were easier for it to reason about than the official, correct schematics. A representative example is shown in Table \ref{tab:qwq-example}. In this example, the main difference lies in how the 74HC595's DS, SHCP, and STCP pins are connected. In the correct schematics, since some tasks involve more complex hardware setups with limited digital ports, we standardize the pin connections to A0, A1, and A2 on the Arduino. However, for this specific task, connecting to pins 2, 3, and 4 would be a more natural and efficient choice. The LLM selected this alternative schematics in the ``Setting \ding{203}'' setting, leading to a smoother reasoning process. This small change had a significant impact. In the ``Setting \ding{203}'' task, QWQ used only 1,565 tokens to construct a digital logic table. In contrast, for the ``Setting \ding{202}'' task—despite being given the correct schematics—it used 9,276 tokens, indicating a much more complex and less efficient reasoning path. \textcolor{darkgreen}{It is not that the human-written schematics are bad or incorrect, but rather that LLMs tend to find it easier to understand their own generated schematics.}

\begin{center}
  \includegraphics[width=\linewidth]{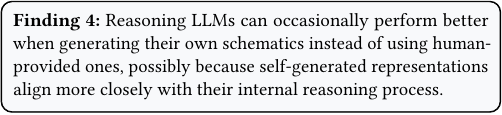}
\end{center}

\subsection{\textcolor{darkgreen}{Techniques to Improve LLMs’ Capability in Embedded System Development}}
\textcolor{darkgreen}{Based on the above analysis, we propose two techniques to improve LLMs’ capability in embedded system development.}

\noindent{\textbf{R1-Retrieval}} As discussed in Section 5.2.2, we observe that LLMs encounter different challenges in the code generation task (Setting \ding{202} and \ding{203}). Chat LLMs can leverage pretrained knowledge but often struggle to adapt flexibly to specific problem contexts. In contrast, reasoning LLMs tend to approach tasks from an overly low-level perspective, missing higher-level abstractions. Ideally, an effective LLM should combine pretrained knowledge with adaptive reasoning capabilities, enabling it to adjust and apply its understanding to the specific problem dynamically.

To address this, we propose a method called R1-Retrieval, which combines the strengths of both DeepSeek-V3 and DeepSeek-R1. The approach is as follows: we first use LLM DeepSeek-V3 to generate code for embedded system tasks that can be addressed using pretrained knowledge alone. This generated code is then provided to DeepSeek-R1 as a form of retrieval-augmented input within its prompt. With this setup, DeepSeek-R1 can leverage the pretrained output of DeepSeek-V3 as contextual knowledge, enabling it to reason more effectively and solve problems that require both learned knowledge and adaptive reasoning ability.

For evaluation, we selected a simple task for each of the seven main electronic components—tasks that DeepSeek-V3 can handle with ease. For each target task, if it involves a particular electronic component, the corresponding code from DeepSeek-V3 is included in the retrieval input. These code and the associated prompts are detailed in the Appendix H 
in \cite{artifact}. The results are summarized in Table \ref{tab:activateResult}. With the R1-Retrieval method, the pass@1 score of DeepSeek-R1 improved from 55.6 to 65.1 on the ``Setting \ding{202}'' task, and from 50.0 to 53.1 on the ``Setting \ding{203}'' task.

\noindent\textbf{R1-Compiler.} For Setting \ding{204} task development using the ESP32 and the ESP-IDF framework, LLMs often generate code with syntax errors, making deployment infeasible. To mitigate this, we introduce R1-Compiler, a method that incorporates the compiler as an external feedback mechanism to guide code repair. Specifically, we capture compiler error messages and feed them back into DeepSeek-R1, prompting it to revise its output accordingly. The detailed prompting strategy is provided in the Appendix I
in \cite{artifact}.

The incorporation of compiler feedback significantly reduces syntax errors from 34.1\% (43/126) to 3.2\% (4/126) after just one repair iteration. The final result is shown in Table \ref{tab:activateResult}, the pass@1 score increases from 21.4 to 27.8.

\subsection{Implications}
Although LLMs perform poorly on our fundamental embedded systems design benchmark, they demonstrate significant potential. Even without correct schematics, they can leverage inherent reasoning abilities to derive improved solutions. We believe that with more data and advanced training paradigms, LLMs could eventually enable interaction between virtual and physical worlds, representing an initial step toward embodied intelligence.

\begin{table}[t!]
  \small
  \resizebox{1.0\linewidth}{!}{
    \begin{tabular}{l|ccc}
    \toprule
    Methods & Setting \ding{202} & Setting \ding{203} & Setting \ding{204} (ESP32) \\
    \midrule
    DeepSeek-R1 & 55.6  & 50.0    & 21.4 \\
    R1-Retrieval & 65.1  & 53.1  & \textbackslash{} \\
    R1-Compiler & \textbackslash{} & \textbackslash{} & 27.8 \\
    \bottomrule 
    \end{tabular}%
    }
    \caption{The result of R1-Retrieval and R1-Compiler}
  \label{tab:activateResult}%
\end{table}%
\section{Related Work}

\noindent \textbf{Benchmark for Embedded System}. 
There are some benchmarks related to embedded systems~\cite{englhardt2024exploring,yang2024embedgenius,liu2023verilogeval}, but \textcolor{darkgreen}{they restrict LLMs to code generation only, whereas our framework supports full end-to-end embedded system design. Furthermore, our validation pipeline is fundamentally different and more rigorous. Prior benchmarks typically inserted print statements into the generated code and determined correctness from the resulting logs, a strategy that models could easily exploit. In contrast, our approach conducts precise, automated, end-to-end verification, eliminating such shortcuts and ensuring a reliable assessment of model performance.} Evaluating LLMs on real hardware is also significantly more costly and resource-intensive. Our benchmark leverages the Wokwi simulation platform, enabling automated evaluation that closely reflects real hardware behavior, without the time and resource costs associated with using physical hardware components.

\noindent \textbf{Large Language Models for Embedded Systems}. 
Existing approaches that use LLMs for hardware platform programming primarily rely on integrated development environments or programming frameworks. However, they rarely enable genuine interaction between the LLM and the hardware\cite{ball2019microsoft,brennan2022exploring,devine2019makecode,koopman2005undergraduate,pasricha2022embedded}. Some studies investigated the use of LLMs for hardware-related work. These include generating code for Field Programmable Gate Arrays (FPGAs)\cite{guo2008efficient,podili2017fast,moreira2010automatic}, using LLMs to write Arduino code but focusing on human-in-the-loop testing\cite{englhardt2024exploring}, or leveraging serial output to provide additional information for LLMs debugging\cite{yang2024embedgenius}. Despite these advancements, most of the work remains limited to code generation and does not involve complete embedded system design. Our benchmark is the first to allow an LLM to independently implement the full process of embedded system development.
\section{Threat to Validity}
Our study identifies four potential threats to validity, along with the steps we have taken to mitigate them. 

First, the manual construction of our benchmark dataset introduces the possibility of subjective bias in labeling. 
To reduce this risk, two independent annotators labeled the data separately, and any differences were discussed and resolved through consensus. 

Second, the LLMs used in our experiments inherently involve randomness, which can cause variability in the observed results. To mitigate this issue, we set the decoding temperature to 0 and employed greedy decoding during inference, ensuring deterministic and consistent results across multiple runs. 

Third, our evaluation relies on the correctness and stability of the Wokwi simulation platform. Wokwi supports a large and active user community, hosting more than one million Arduino projects, 800,000 ESP32 projects, and 100,000 Raspberry Pi Pico projects. \textcolor{darkgreen}{While alternative simulators such as Renode and SimulIDE are available, we found them less convenient and harder to integrate into automated evaluation pipelines.} During the process of creating our benchmark and associated test cases, we encountered no unexpected behaviors, which enhances our confidence in the platform's reliability.

\textcolor{darkgreen}{Fourth, there may exist hardware-specific issues, but they are not the central challenge this paper addresses. Our core contribution is to provide a reproducible benchmark for evaluating the reasoning and design capabilities of LLMs. A simulation platform is the ideal tool for this purpose. It offers a controlled and fully automated environment that eliminates external noise, making our evaluation both reliable and scalable. This approach is not only consistent with prior work~\citep{widianto2023utilization} but also fundamentally necessary for a robust and fair assessment of LLM performance.}

\section{Conclusion}
In this work, we propose a benchmark to evaluate the ability of LLMs in embedded system design. Our benchmark consists of four types of tasks, each with 126 cases. These tasks assess the LLMs' capabilities in embedded programming with given wiring, autonomous circuit wiring, end-to-end embedded system design, and cross-platform migration. We also provide an automated and accurate evaluation method and conduct experiments on 10 state-of-the-art LLMs. Through our analysis, we find that although the LLMs face challenges in solving these tasks, they also demonstrate significant potential. For instance, discovering wiring schemes that are even better than the reference solutions. Based on our findings, we propose several simple strategies that can effectively improve the performance of LLMs in embedded system development.

\begin{acks}
We sincerely thank the reviewers for their insightful comments and valuable suggestions. This work was supported by the Natural Science Foundation of China (No.  62572456, 62272439, 62306303), CAS Project for Young Scientists in Basic Research (Grant No.YSBR-040), Hong Kong SAR Research Grant Council (General Research Fund Number 16206524).
\end{acks}
\newpage

\bibliographystyle{ACM-Reference-Format}
\bibliography{sample-base}

\end{document}